\documentclass{mem}
\usepackage{natbib}\usepackage{txfonts}\usepackage{balance}
\usepackage{graphicx}
\usepackage{txfonts}
\usepackage[a4paper]{hyperref}
\idline{79}{3}
\begin{document}

\title{Detection of new variable stars in the SMC cluster NGC 121
}

   \subtitle{}

\author{
G. \,Fiorentino\inst{1,2} \and R. \, Contreras\inst{1} \and G. \,
Clementini\inst{1} \and K. Glatt\inst{3} \and E. \, Sabbi\inst{4}
\and M. \, Sirianni\inst{4} \and E. \, Grebel\inst{3} \and J. \,
Ghallager\inst{5}
          }

  \offprints{G. Fiorentino}

\institute{
Istituto Nazionale di Astrofisica --
Osservatorio Astronomico di Bologna, Via Ranzani 1,
I-40127 Bologna, Italy
\email{giuliana.fiorentino@inaf.oabo.it}
\and
Kapteyn Astronomical Institute, University of Groningen, Postbus 800, 9700AV Groningen, the Netherlands
\and
University of Basel, Venusstrasse 7, CH-4102 Binningen, Switzerland
\and
Space Telescope Science Institute, 3700 San Martin Drive, Baltimore,
MD 21218, US
\and
University of Wisconsin, 55343 Sterling Hall, 475 North Charter Street, Madison, WI 53706-1582
}

\authorrunning{Fiorentino et al.}

\titlerunning{RR Lyrae in NGC121}

\abstract{New candidate variable stars have been identified in the
Small Magellanic Cloud cluster NGC121, by applying both the image
subtraction technique (ISIS, Alard 2000) and the Welch \& Stetson
(1993) detection method to HST WFPC2 archive and ACS proprietary
images of the cluster. The new candidate variable stars are located from the
cluster's Main Sequence up to Red Giant Branch. Twenty-seven of
them fall on the cluster Horizontal Branch and are very likely RR
Lyrae stars. They include the few RR Lyrae stars already
discussed by Walker \& Mack (1988). We also detected 20 Dwarf
Cepheid candidates in the central region of NGC121. Our results
confirm the ``true" globular cluster nature of NGC121, a cluster
that is at the young end of the Galactic globulars' age range.
} \maketitle{}

\section{Introduction}
NGC121 is a key cluster to understand the Star Formation History
(SFH) occurred in the SMC through the study of the similarities
and differences with respect to the Milky Way clusters. In fact,
with an estimated age of about 10 Gyrs, it locates at the
transition between open and globular clusters in the MW. Moreover,
as emphasized by Stryker et al. (1987), NGC121 very likely marks
the boundary between ``old'' Population II (containing RR Lyrae
but not carbon stars) and intermediate-age populations (containing
carbon stars but not RR Lyrae stars). Up to now, only four RR
Lyrae stars with light curves derived from ground-based
observations, were known in NGC121 (Walker \& Mack 1988), and
references therein). As part of a coordinated HST and ground-based
effort, we have observed the SMC with the ACS (Cycle13, GO
prog.10396, PI Gallagher), pointing at 7 star clusters of
different age and metallicity (including NGC121) and 7 fields in
various galactic locations (center, periphery, wing and bridge
towards the LMC), to derive their key evolution parameters and SF
histories (see e.g. Glatt et al. 2007, submitted). The ACS
observations of NGC121 were taken in time-series fashion in order
to use them, along with the existing WFPC2 archive data of the
cluster, to detect new variable stars in the cluster.

\section{Observations and data reductions}

\begin{figure}
\includegraphics[width=7cm]{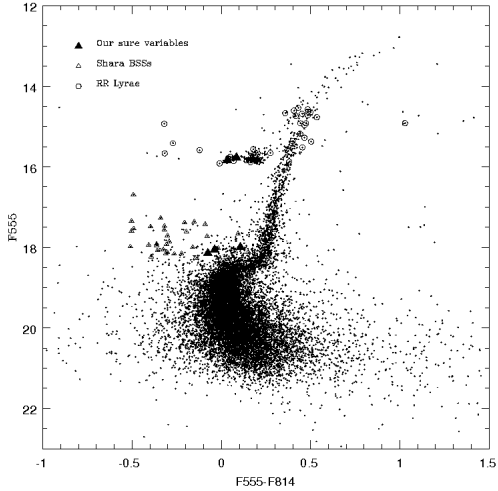}
\caption[]{CMD of NGC 121 for stars on the Planetary Camera (PC)
of the WFPC2. Variable stars are marked by different symbols. {\it
Open circles}: RR Lyrae stars; {\it Open triangles}: Dwarf
Cepheids; {\it Filled triangles}: Sure variables.}
\end{figure}

The dataset used for our analysis consists of 18 images of NGC121
in the $V$ (F555W) band and 12 images in the $I$ (F814W) band,
taken with the Wide Field Planetary Camera 2 (WFPC2, Dolphin et
al. 2001) and with both the High-Resolution Camera (HRC) and the
Wide Field Camera (WFC) of the Advanced Camera for Surveys (ACS)
on board of the HST. The images cover temporal ranges of 2.5
(WFPC2), 0.35 (HRC) and 1.2 (WFC) hours in length, respectively.
Since the primary goals of both proprietary and archive
observations were to derive  key stellar population parameters for
the SMC, from the analysis of deep $V, I$ color-magnitude
diagrams, the observations generally consisted of a few deep
images obtained with long exposure times. For these reasons our
images, while still adequate to detect variable sources, are not
adequate to derive periodicities for variables of RR Lyrae type,
whose typical periods are in the range from 7 to 12-15 hours.
Nevertheless, we succeeded to identify and build portions of the
light curve for many new RR Lyrae star candidates, and we also
detected for the first time several Dwarf Cepheids (DCs) in
NGC121. The first data reduction was performed with the
DAOPHOT/ALLSTAR/ALLFRAME package (Stetson 1987, 1994), but our
best and deepest photometry was obtained with the Dolphot/HSTphot
packages, implemented by Dolphin (2000, and reference therein)
``ad hoc" to perform the photometry of WFPC2 and ACS images,
respectively. An additional advantage of the Dolphot/HSTphot
packages is to output the star magnitudes already calibrated in
the Johnson-Cousins photometric system, making easier the study of
the light curves of the candidate variables.

\section{Variable star identification \& conclusions}

\begin{figure}
\includegraphics[width=6cm]{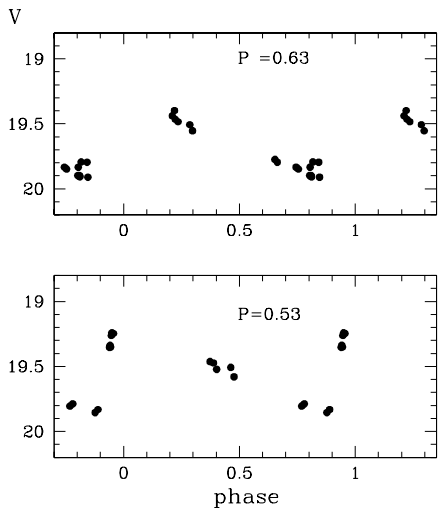} \caption[]{Light curves of fundamental-mode
RR Lyrae stars. {\it Upper panel}: new RR Lyrae star with $P \sim$
0.63 days. {\it Lower panel}: the fundamental-mode RR Lyrae star
V37 discovered by Walker \& Mack (1988).} \label{map}
\end{figure}

The NGC121 candidate variable stars were identified by applying
both the image subtraction technique (ISIS, Alard 2000) and the
Welch \& Stetson (1993) detection method. We identified about 50
candidate variables in NGC121. Period search was performed using
GRaTiS (Graphical Analyzer of Time Series) a private software
developed at the Bologna Observatory (see e.g. Clementini et al.
2000). Twenty-seven of the candidate variables are located on the
cluster Horizontal Branch, thus are very likely RR Lyrae stars.
They include the few RR Lyrae stars already discussed by Walker \&
Mack (1988). We also detected 20 Dwarf Cepheid candidates in the
central region of NGC121, and recovered 34 Blue Straggler star
candidates found by Shara et al. (1998). DCs are late-A and
early-F type stars, that populate the instability strip near or
slightly above the zero-age main sequence and have magnitude from
0.2 to about 3 mag fainter than the RR Lyrae stars. In the CMD
they fall where the instability strip crosses the region of the
Blue Straggler stars. These variables have typical periods in the
range from 1 to 6 hours, thus our data sampling allow to sample
their light curves better than for the RR Lyrae stars.

\begin{figure}
\includegraphics[width=6cm]{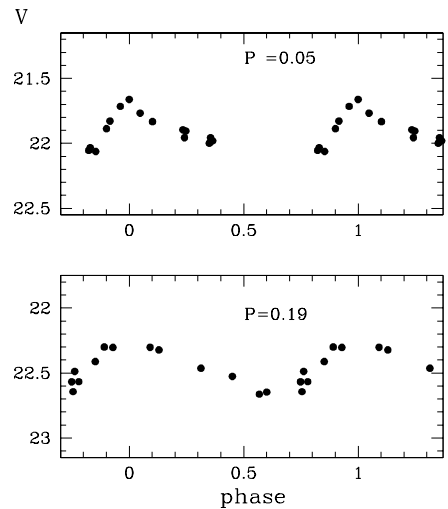} \caption[]{Light curves of two Dwarf Cepheids.
{\it Upper panel}: SX Phoenicis star with $P$=0.05 days. {\it
Lower panel}: $\delta$ Scuti star with $P$=0.19 days. }
\end{figure}

DCs are divided into: 1) $\delta$ Scuti stars, which are metal
rich, young, Population I stars with typical periods longer than
0.1 day usually observed in open clusters and in the field of the
MW; and 2) SX Phoenicis stars, which are metal poor, Population II
variables, generally observed in GCs, with typical periods in the
range from 0.03 to 0.09 days (Poretti et al. 2006). NGC121 seems
to host both types of DCs. Fig. 1 shows the CMD of NGC121 for
stars on the PC of the WFPC2, obtained using DAOPHOT/ALLFRAME for
the data reduction. The variable stars are marked by different
symbols (see caption). Examples of light curves for
RR Lyrae stars and DCs are shown in Figs. 2-3.\\

The conspicuous number of RR Lyrae stars detected in NGC121
confirms the ``true" globular cluster nature of NGC121, a cluster
that is at the young end of the Galactic globulars' age range. The
apparent presence of both $\delta$ Scuti and SX Phoenicis stars in
NGC121 is an intriguing feature, that needs to be confirmed by a
more detailed analysis of the light curves and a careful study of
the contaminations by the SMC field stars.

\bibliographystyle{aa}

\end{document}